\documentclass[3p,12pt]{elsarticle}

\usepackage[colorlinks=true,citecolor=blue,linkcolor=blue]{hyperref}
\usepackage{amsmath}
\usepackage{amssymb}
\usepackage{color}
\usepackage{graphicx}
\usepackage{longtable}
\usepackage{listings}

\usepackage{chngcntr}
\counterwithin{figure}{section}

\newcommand{\bra}[1]{\ensuremath{\left\langle#1\right|}}
\newcommand{\ket}[1]{\ensuremath{\left|#1\right\rangle}}
\newcommand{\bracket}[2]{\ensuremath{\left\langle#1 \vphantom{#2}\right| \left. #2 \vphantom{#1}\right\rangle}}

\newcommand{\bls}{\begin{lstlisting}}
\newcommand{\els}{\end{lstlisting}}

\usepackage{xparse}
\ExplSyntaxOn
\NewDocumentCommand{\mref}{m}{\quinn_mref:n {#1}}
\seq_new:N \l_quinn_mref_seq
\cs_new:Npn \quinn_mref:n #1
 {
  \seq_set_split:Nnn \l_quinn_mref_seq { , } { #1 }
  \seq_pop_right:NN \l_quinn_mref_seq \l_tmpa_tl
  ( % print the left parenthesis
  \seq_map_inline:Nn \l_quinn_mref_seq
    { \ref{##1},\nobreakspace } % print the first references
  \exp_args:NV \ref \l_tmpa_tl % print the last or only one
  ) % print the right parenthesis
 }
\ExplSyntaxOff

\newcommand{\pa}[2]{\frac{\partial #1}{\partial #2}}

\definecolor{mauve}{rgb}{0.88, 0.69, 1.0}
\definecolor{dkgreen}{rgb}{0,0.6,0}

\begin{document}

\bibliographystyle{unsrt}
\begin{frontmatter}
\title{A Simple Quantum Integro-Differential Solver (SQuIDS)}

\author[phy,WIPAC]{Carlos A. Arg\"uelles Delgado}
\ead{carlos.arguelles@icecube.wisc.edu}
\author[phy,WIPAC]{Jordi Salvado }
\ead{jsalvado@icecube.wisc.edu }
\author[phy,WIPAC]{Christopher N. Weaver}
\ead{cweaver@icecube.wisc.edu}

\address[phy]{Department of Physics, University of Wisconsin \\ Madison,
  WI 53706, USA} 
\address[WIPAC]{Wisconsin IceCube Particle Astrophysics Center\\ Madison, WI 53706, USA} 

\tnotetext[t1]{The code can be found in \url{https://github.com/jsalvado/SQuIDS}}

\journal{Computer Physics Communications}

\begin{abstract}
Simple Quantum Integro-Differential Solver (SQuIDS) is a C++ code
designed to solve semi-analytically the evolution of a set of density
matrices and scalar functions. This is done efficiently by expressing all operators in an SU(N) basis. 
SQuIDS provides a base class from which users can derive new classes
to include new non-trivial terms from the right hand sides of density matrix equations.
The code was designed in the context of solving neutrino oscillation problems,
but can be applied to any problem that involves solving the quantum evolution
of a collection of particles with Hilbert space of dimension up to six.
\end{abstract}
\end{frontmatter}

\hypersetup{linkcolor=black}
\tableofcontents
\hypersetup{linkcolor=blue}

\section{Introduction}
\label{intro}
The evolution of an ensemble of neutrinos is a many body quantum mechanics 
problem, where every neutrino is represented by a state in a Hilbert space.

When all neutrinos are produced in the same quantum state and
there are no interactions, a pure quantum state evolution is a good 
approximation. This can be implemented by solving the corresponding Schr\"odinger equation.

On the other hand, especially at high energies, there are interactions
which can mix the states. In this case, the system is represented by a
mixed state and therefore the evolution is more naturally described in
the context of the density matrix formalism
\cite{GonzalezGarcia:2005xw,Wolfenstein:1977ue,Sigl:1992fn}.

The possibility of using the symmetries of the problem in order to efficiently solve
the evolution, i.e. differential equations, is a well studied topic in applied
Mathematics \cite{LIE}. In this paper, we apply this principle to
make explicit the physical degrees of freedom, thus enabling us to solve 
analytically the evolution due to the time independent part of the Hamiltonian ($H_0$). 
This approach has been also used in the context of two neutrino flavor approximations \cite{Sigl:1992fn}.
 
The Simple Quantum Integro-Differential Solver (SQuIDS) is a code that
implements the evolution of a set of density matrices and scalar functions using a 
semi-analytic approach.
The code is written in object-oriented C++ and contains a base class which sets up the 
problem by means of virtual member functions. This allows the user to create a derived class which defines the
right hand side terms of the differential equation via its implementation of these functions.
Numerical integration is performed using the GNU Scientific Library \cite{GSL}.

The code works with a Hilbert space of dimension up to six and allows the
inclusion of an arbitrary number of density matrices and scalar functions.
The physical degrees of freedom in the problem can be represented in the basis of the generators
of the $SU(N)$ group plus the identity, where $N$ is the dimension of
the Hilbert space, so we write all operators in terms of this basis. 
The evolution generated by the time independent part of 
the Hamiltonian and basis changes can be thought as the
action of the unitary group on the operators. These
transformations are included as analytic expressions in the code.
Then, in order to make the numerical solution efficient we solve the
evolution of the entire system in the interaction picture \cite{Arguelles:2012cf}.
 
The paper is structured in the following sections: In section \ref{DMF}
we introduce the density matrix formalism. 
In section \ref{previous}, we comment on other approaches to solve similar problems. 
In section \ref{SUNDMF}, we describe the
evolution equations in the context of the $SU(N)$ algebra.
In section \ref{Descript}, we describe how this formulation is implemented
in the code.
Finally, in section \ref{Examples}, we include 
three simple examples to illustrate some applications: neutrino oscillations in vacuum, Rabi oscillations, and collective neutrino oscillations.

\section{Related work}
\label{previous}

Solving quantum mechanics problems using computers has been a long standing topic of study. Examples
of codes that solve the Schr\"{o}dinger equation in the position representation have been published for 
both stationary \cite{Wensch2004129} and time evolving \cite{MoxleyIII20122434,Dziubak2012800,Marques200360} many body systems.
On the other hand, representation of finite, quantum-mechanical, closed systems has been developed in order to 
perform quantum computing calculations (see e.g. \cite{libquantum}). In high energy particle physics
the need for precise neutrino oscillation calculation has encouraged the development of ad hoc tools to solve two or three level
closed quantum systems subject to time varying potentials (see e.g. \cite{prob3++,Wallraff:2014vl,Huber:2007ji}). This work goes beyond the pure state representation by using the density matrix formalism, which allows us to express in a natural way mixed states, as well as treating open quantum systems. To do this, we have developed a highly efficient representation of states and operators in terms of the generators of the SU(N) group. 
  
\section{Density matrix formalism}
\label{DMF}

\subsection{Definition}

In quantum mechanics the state of system is given by a vector in a
Hilbert space, i.e.
$\ket{\phi_i} \in \mathcal{H}_{i}$, where $\mathcal{H}_i$ is the
Hilbert space for the $i$-th particle. In general we can be interested
in solving a system of many particles and for that the Hilbert space
is constructed as $\mathcal{H}= \bigotimes_i \mathcal{H}_i$. For a
large number of particles the dimension of this space grows
exponentially. 
Nevertheless in the limit where the different particles do not
have quantum correlations and $\mathcal{H}_i$ is the same for all of
the single particles of the system, we can approximate the system as a 
statistical ensemble of single particle quantum states, which is known
as a mixed state.

To describe a system in this limit it is convenient to
introduce the density matrix formalism. For a given set of
states $\{\bra{\psi_i}| i=1,...,n\}$ and a set of positive real
numbers $\{p_i | \sum_i p_i=1  \}$, that are physically interpreted as the
probability of the system to be in the $i$-th state, the density
operator, which represents the mixed state, is constructed as 

\begin{equation}
\hat\rho=\sum_i p_i \ket{\psi_i}\bra{\psi_i}.
\end{equation}

For a particular basis $\{\ket{u_1},\ket{u_2} ... \ket{u_n} \}$ of the
Hilbert space the components of the density matrix can be written as 

\begin{equation}
\label{eq:rhodeff}
\rho_{lm}=\sum_i p_i \bracket{u_l}{\psi_i} \bracket{\psi_i}{u_m},
\end{equation}
in other words, for a given basis the density operator is represented by a $n
\times n $ Hermitian matrix $\rho$.

Is also useful to define the projectors to the $i$-state of a basis,

\begin{equation}
\label{eq:proj}
\hat\Pi^{(i)}=\ket{u_i}\bra{u_i}.
\end{equation}

The expectation value for any operator $\hat O$ when the system is described by
$\hat\rho$ is given by the trace of the matrix product

\begin{equation}
\label{eq:expected_val}
\left\langle \hat O \right\rangle_\rho =\sum_{lm} O_{lm}\rho_{lm},
\end{equation}
where $O_{lm}$ are the components that represent $\hat O$ in the same 
basis as $\rho_{lm}$.

In particular, the expectation value of the projector gives the
probability of finding the system in the $i$-state, and is

\begin{equation}
\label{eq:expvalproj}
\left\langle \hat\Pi^{(i)} \right\rangle_\rho=\sum_{lm}\Pi^{(i)}_{lm}\rho_{lm}=p_i.
\end{equation}

\subsection{Unitary transformations}
\label{UnitaryTrans}

In general in quantum mechanics, the transformation from one basis to
another is always
given by a unitary transformation. In this section we describe a
way of writing a general unitary transformation in terms of a set
of mixing angles and phases.
This parametrization is widely used in the context of neutrino physics 
and is included in the code.

This parametrization can be written as product of two dimensional 
rotations, in particular, 

\begin{equation}
\label{eq:mixing}
U(\theta_{ij}\delta_{ij})=R_{N-1\, N} R_{N-2\, N} ... R_{45} R_{35} R_{25} R_{15}
R_{34} R_{24} R_{14} R_{23} R_{13} R_{12}, 
\end{equation}
where each matrix $R_{ij}$ is a rotation in the $ij$ plane with rotation 
angle $\theta_{ij}$ and complex phase $\delta_{ij}$, namely

\begin{align}
\label{eq:mixingdeff}
  R_{ij} = \begin{pmatrix}
              \ddots & \\
                     & \cos\theta_{ij}  & \cdots & \sin\theta_{ij} e^{-i \delta_{ij}}\\
                     & \vdots           &        & \vdots \\
                     & -\sin\theta_{ij} e^{i \delta_{ij}} & \cdots & \cos\theta_{ij} \\
                    &                  &        &                 & \ddots
           \end{pmatrix} \,.
\end{align}
In a minimal description only a subset of the $\delta_{ij}$ are required to be allowed to be non-zero. 

An operator in the Hilbert space is transformed using the relation

\begin{equation}
\label{eq:operator_rot}
O_{lm} \rightarrow \sum_{kn} U_{lk}^\dagger O_{kn} U_{nm}.
\end{equation}

\subsection{System Definition and Time Evolution}
\label{sec:system_time_def}

Consider a system corresponding to $n_x$ nodes, where a node is an object with $n_\rho$ 
density matrices and $n_s$ scalar functions, labeled by a parameter $x$. 
For notational convenience we define  $\tilde\rho$ and $\tilde S$ as the set of all density
matrices with elements  $\rho_J$ and set of all scalar functions with elements $S_K$, respectively, where $J$
and $K$ are general indices to label the elements of the sets, 
$J \in \{ (i,j) | i=0, ...,n_x-1  ; j=0, ...,n_\rho-1\}$ and
$K \in \{ (i,k) | i=0, ...,n_x-1  ; k=0, ...,n_s-1\}$.

The evolution of the system is given by two coupled equations: one for the density matrices 
and another for the scalar functions. The first one is the Von Neumann
equation which contains terms for coherent, decoherence, and
other interactions. The second one is a Boltzmann-like differential equation for
the scalar functions

\begin{eqnarray}
\label{eq:rhoevol}
\pa{\rho_J}{t} &=& -i[H_J(t),\rho_J]+ \{\Gamma_J(t),\rho_J\}+ 
F_J(\tilde\rho,\tilde S;t) \;,\\ 
\label{eq:rhoevol-2}
\pa{S_K}{t} &=& -\Gamma_K(t) S_K+G_K(\tilde\rho,\tilde S;t) \;,
\end{eqnarray}
where $H_J$ is the Hamiltonian, $\Gamma_J$ is the decoherence and
attenuation term, $\Gamma_K$ is the attenuation for the scalar
functions, and finally the functions $F_J$ and $G_K$ are general
functions of the full sets $\tilde\rho$ and $\tilde S$ that may contain non-linear
and interaction terms.

The scalar functions can be used to solve the evolution of populations
of particles where there is no coherent quantum dynamics. 
These parameters can also be used to implement any other first-order differential equation. 

\subsubsection{Interaction Picture Evolution}

The Hamiltonian can always be decomposed into one part that does not
depend on time $H_{0J}$ and another that contains the time dependence,
$H_{1J}(t)$

\begin{equation}
H_J(t)=H_{0J}+H_{1J}(t) \;.
\end{equation}

In a finite Hilbert space the evolution generated by $H_{0J}$
can be solved analytically. This may dramatically speed up the numerical
computation, particularly in the case where $H_0$ contains terms that
produce very fast oscillations.
This motivates expressing the problem in the interaction picture in which 
an operator $\hat O_J$ is transformed into $\bar O_J$ by

\begin{equation}
\label{eq:time}
\bar{O}_J(t)=e^{-iH_{0J}t}\hat{O}_Je^{iH_{0J}t}.
\end{equation}

Then, the evolution equations Eq.\eqref{eq:rhoevol} are

\begin{eqnarray}
\label{eq:rhoevol2}
\pa{\bar\rho_J}{t} &=& -i[\bar H_{1J}(t),\bar \rho_J]+ \{\bar\Gamma_J(t),\bar\rho_J\}+ 
F'_J(\tilde{\bar \rho},\tilde S;t) \;,\\ 
\pa{S_K}{t} &=& -\Gamma_K(t) S_K+G'_K(\tilde{\bar \rho},\tilde S;t) \;,
\end{eqnarray}
where $\bar H_{1J}(t)$ is known as the interaction Hamiltonian, and $F'_J$ and $G'_K$ are 
the corresponding $F_J$ and $G_K$ in the interaction picture.

\section{Density matrix formalism using $SU(N)$ generators}
\label{SUNDMF} 
Every $\rho_J$ is a Hermitian matrix therefore has, by construction, real eigenvalues 
and can always be decomposed as a
linear combination of the $SU(N)$ generators plus
the identity with real coefficients, being $N$ the dimension of the Hilbert space. The
same applies to all the operators in the right hand side of Eq.\eqref{eq:rhoevol}.
Using this decomposition we can write the operators as follows,

\begin{eqnarray}
\label{eq:SUN_decomp}
\rho_J &=& \rho_J^\alpha \lambda_\alpha, \\
H_J &=&  H_J^\alpha \lambda_\alpha, \\
\Gamma_J &=& \Gamma_J^\alpha \lambda_\alpha, 
\end{eqnarray}
where $\{\lambda_\alpha | \alpha=1,...,N^2-1\}$ are the $SU(N)$
generators,  $\lambda_0=\mathbb{I}$, and  we have use the Einstein convention, i.e.
a sum over repeated indices is implicit.

We can use the commutator and anti-commutator relations in the
lie algebra,

\begin{eqnarray}
\label{eq:SUN_alg}
i[\lambda_\alpha, \lambda_\beta] &=& - f^\gamma_{\alpha\beta
  } \lambda_\gamma \\
  \label{eq:SUN_alg-2}
\{\lambda_\alpha, \lambda_\beta\} &=& d^\gamma_{\alpha \beta
  } \lambda_\gamma \;,
\end{eqnarray}
in order to write down the evolution equations Eq.\mref{eq:rhoevol,eq:rhoevol-2}
\begin{eqnarray}
\label{eq:rhoevol_sun}
\pa{\rho^\alpha_J}{t} &=& H^\beta_J(t)\rho^\gamma_J
f^\alpha_{\beta\gamma}+ \Gamma^\beta_J(t)\rho^\gamma_J
d^\alpha_{\beta \gamma} + F^\alpha_J(\tilde\rho,\tilde S;t) \;,\\ 
\pa{S_K}{t} &=& -\Gamma_K(t) S_K+G_K(\tilde\rho,\tilde S;t) \;.
\end{eqnarray}

\subsection{Unitary transformations and zero order time evolution}
\label{UnitTrans}
Since any hermitian operator $\hat O$ can be expressed as linear combination of 
$\{\lambda_\alpha\}$, then the effect of any linear transformation on $\hat O$ only
depends on the effect on every element on the basis $\{\lambda_\alpha\}$. 
In particular, transformations of the form given in Eqs.\mref{eq:mixing,eq:time} 
fall into this category. For Eq.\eqref{eq:mixing} we can write this explicitly
\begin{equation}
\label{eq:trans1}
\lambda_\alpha \rightarrow  U^\dagger(\theta_{ij}\delta_{ij}) \lambda_{\alpha} U(\theta_{ij}\delta_{ij}),
\end{equation}  
given that Eq.\eqref{eq:trans1} preserves hermiticity and that $\{\lambda_\alpha\}$ 
spans the hermitian operator space then

\begin{equation}
  \label{eq:rotequation}
  U^\dagger(\theta_{ij}\delta_{ij})\lambda_{\alpha }
  U(\theta_{ij}\delta_{ij}) =  u^\beta_\alpha(\theta_{ij}\delta_{ij}) \lambda_\beta,
\end{equation}  
where $u^\beta_\alpha(\theta_{ij}\delta_{ij})$ are real functions that are solved analytically in 
the code.

The same procedure can be applied to the time evolution generated by $H_0$
given by Eq.\eqref{eq:time}, i.e.

\begin{equation}
  \label{eq:timeequation}
  \bar\lambda_\alpha(t) = e^{-iH_0t} \lambda_{\alpha } e^{iH_0t}
  =  u^\beta_\alpha(t) \lambda_\beta,
\end{equation}  
as before the analytic expressions for $u^\beta_\alpha(t)$ are in the code.

\subsection{Interaction Picture Evolution}

Give the formalism in Sec. \ref{UnitTrans} a general hermitian operator $\hat O$ in the 
interaction picture is given by

\begin{equation}
\bar{O}(t)=e^{-iH_0t}O e^{iH_0t}=O^\alpha e^{-iH_0t}\lambda_\alpha
e^{iH_0t} = O^\alpha \bar\lambda_\alpha(t) \;. 
\end{equation}

The commutator and anti-commutator relations holds for the time
dependent generators with the same structure constants, which implies
that the evolution equations in the interactions picture are the same,
but with the $\bar H_{1J}(t)$ instead of $H_{J}(t)$, 

\begin{eqnarray}
\label{eq:rhoevol_sun_int}
\pa{\bar \rho^\alpha_J}{t} &=& \bar H^\beta_{1J}(t)\bar \rho^\gamma_J
f^\alpha_{\beta\gamma}+ \Gamma^\beta_J(t)\bar \rho^\gamma_J
d^\alpha_{\beta \gamma} + F'^\alpha_J(\tilde{\bar \rho},\tilde S,t) \;,\\ 
\label{eq:rhoevol_sun_int-2}
\pa{S_K}{t} &=& -\Gamma_K(t) S_K+G'_K(\tilde{\bar \rho},\tilde S,t) \;.
\end{eqnarray}

\section{Description of the Code}
\label{Descript} 

SQuIDS is a code written in C++ to solve the set of equations specified in
the previous section, specifically Eqs.\mref{eq:rhoevol_sun_int,eq:rhoevol_sun_int-2}. In this section 
we will describe the classes, functions, and operations defined between the objects 
that compose the SQuIDS library.

The system structure is shown in Figure \ref{fig:squids}, each $i$-node contains a set of density matrices $\{\rho_{(ij)}\}$ and scalar functions $\{S_{(ik)}\}$
as well as a real number $x_i$ that can play the role of a parameter or label for the node. For example, when we are describing
a system in which each node represents a given position $x$ then $x_i \equiv x$, on the other hand when we are describing a system in which
each state its characterized by is energy or momentum, then  $x_i \equiv E$ in each node. Even though $x_i$ can have a physical interpretation, in general it need not to be related to any physical quantity. Furthermore, each node can have an arbitrary number of scalars and density matrices, according to the problem at hand.

All of the classes are declared in the header files contained in the folder {\ttfamily  SQuIDS/inc/}, and the corresponding source code 
that implements them is in {\ttfamily  SQuIDS/src/}. In particular, the files that contain the analytic solutions for the evolution 
and rotation Eqs.\mref{eq:rotequation,eq:timeequation} is included in {\ttfamily  SQuIDS/inc/SU\_inc/} up to dimension six. Finally, {\ttfamily  SQuIDS/resources} contains a Mathematica notebook 
that can be used to generate higher dimension solutions.

\subsection{Const}
\label{sec:const}

The {\ttfamily Const} class serves two purposes in the {\ttfamily SQuIDS} library. First, it contains
transformation of physical units to natural units, which enable the user to work on a consistent
unit system, as well as some fundamental constants such as Fermi constant, Avogadro number, proton mass, neutron mass, etc. 
These constant can be access as public members of the class. The full list of constant is listed in Table \ref{tbl:const_prop}. 
Second, it manages the mixing angles ($\theta_{ij}$) and complex phases ($\delta_{ij}$) that define the transformation between the {\ttfamily B0} and {\ttfamily B1} basis
as defined in Eq.\eqref{eq:mixing}, which can be modified and obtained through public member functions. Finally, the class is declared in  {\ttfamily  SQuIDS/inc/const.h} and implemented in {\ttfamily  SQuIDS/inc/const.cpp}.

\subsubsection{Constructors}

\begin{itemize}
  
\item Constructor;
  
  \begin{lstlisting}
    Const();
  \end{lstlisting}
  Initializes all units and arrays.
  
\end{itemize}

\subsubsection{Functions}
  
\begin{itemize}
\item Set Mixing angle  
  \begin{lstlisting}
    void SetMixingAngle(unsigned int i,unsigned int j, 
                        double theta);
  \end{lstlisting}
  
  Sets the rotation angle $\theta_{ij}$ to \texttt{theta}.
\item Get Mixing angle  
  \begin{lstlisting}
    double GetMixingAngle(unsigned int i,unsigned int j);
  \end{lstlisting}
  
  Returns the rotation angle $\theta_{ij}$.
\item Set complex phase  
  \begin{lstlisting}
    void SetPhase(unsigned int i,unsigned int j, double phase);
  \end{lstlisting}
  
  Set the complex phase $\delta_{ij}$ to \texttt{phase}.

\item Get complex phase
  \begin{lstlisting}
    double GetPhase(unsigned int i,unsigned int j);
  \end{lstlisting}
  
  Returns the complex phase $\delta_{ij}$.

\item Set energy eigenvalue difference
  \begin{lstlisting}
    void SetEnergyDifference(unsigned int i, double Ediff);
  \end{lstlisting}
  
  Sets the energy eigenvalue difference $\Delta E_{\mathtt{i}0} = E_{\mathtt{i}} - E_0$ to \texttt{Ediff}. We require $\mathtt{i} > 0$.
  
\item Get energy eigenvalue difference  
  \begin{lstlisting}
    double GetEnergyDifference(unsigned int i);
  \end{lstlisting}
  
  Returns the energy eigenvalue difference $\Delta E_{i0}$.
  
\end{itemize}

\begin{longtable}{|c|p{11cm}|}
      \hline
      %Member & Description \\ \hline \hline
      \hline
      \multicolumn{2}{|c|}{\bf Physical quantities} \\ \hline \hline
      {\ttfamily GF}  & Fermi constant in natural units \\ \hline
      {\ttfamily Na}  & Avogadro number \\ \hline
      {\ttfamily sw\_sq}  & $\sin^2(\theta_w)$ where $\theta_w$ is the weak mixing angle \\ \hline
      {\ttfamily G}  & gravitational constant \\ \hline
      {\ttfamily proton\_mass}  & proton mass\\ \hline
      {\ttfamily neutron\_mass}  & neutron mass \\ \hline
      {\ttfamily electron\_mass}  & $e$ mass  \\ \hline
      {\ttfamily muon\_mass}  & $\mu$ mass  \\ \hline
      {\ttfamily muon\_lifetime}  & $\mu$ lifetime \\ \hline
      {\ttfamily tau\_mass}  & $\tau$ mass  \\ \hline
      {\ttfamily tau\_lifetime}  & $\tau$ lifetime \\ \hline
      {\ttfamily alpha} & fine structure constant \\ \hline
      {\ttfamily e\_charge} & unit of electric charge (Lorentz-Heaviside)\\ \hline
      \hline 
      \multicolumn{2}{|c|}{\bf Units in natural units ($\hbar = c = k_b = 1$)} \\ \hline \hline
      {\ttfamily eV}  & $1$ electron-volt \\ \hline
      {\ttfamily keV}  & $10^3$ electron-volt \\ \hline
      {\ttfamily MeV}  & $10^6$ electron-volt \\ \hline
      {\ttfamily GeV}  & $10^9$ electron-volt \\ \hline
      {\ttfamily TeV}  & $10^{12}$ electron-volt \\ \hline
      {\ttfamily Joule}  & 1 Joule \\ \hline
      {\ttfamily kg}  & 1 kilogram \\ \hline
      {\ttfamily gr}  & 1 gram \\ \hline
      {\ttfamily sec}  & 1 second \\ \hline
      {\ttfamily hour}  & 1 hour \\ \hline
      {\ttfamily day}  & 1 day \\ \hline
      {\ttfamily year}  & 1 year \\ \hline
      {\ttfamily meter}  & 1 meter \\ \hline
      {\ttfamily cm}  & 1 centimeter \\ \hline
      {\ttfamily km}  & 1 kilometer \\ \hline
      {\ttfamily fermi}  & 1 fermi \\ \hline
      {\ttfamily angstrom}  & 1 angstrom \\ \hline
      {\ttfamily AU}  & 1 astronomical unit \\ \hline
      {\ttfamily ly}  & 1 light year \\ \hline
      {\ttfamily parsec}  & 1 parsec \\ \hline
      {\ttfamily picobarn}  & 1 picobarn \\ \hline
      {\ttfamily femtobarn}  & 1 femtobarn \\ \hline
      {\ttfamily Pascal}  & 1 Pascal \\ \hline
      {\ttfamily atm}  & 1 atmosphere \\ \hline
      {\ttfamily Kelvin}  & 1 Kelvin \\ \hline
      {\ttfamily degree}  & 1 degree in radians \\ \hline
      {\ttfamily C}  & 1 Coulomb (Lorentz-Heaviside)\\ \hline
      {\ttfamily A}  & 1 Ampere (Lorentz-Heaviside)\\ \hline
      {\ttfamily T}  & 1 Tesla (Lorentz-Heaviside)\\ \hline
%    \end{tabular}

    \caption{Units and physics constants contained in {\ttfamily Const}.}
    \label{tbl:const_prop}
  \end{longtable}

\subsection{SU\_vector}

The {\ttfamily SU\_vector} class is a type that represents an operator in a N-dimensional Hilbert space as a linear combination of the SU(N) generators
and is the building block of the SQuIDS library. The real coefficients of the generator plus the identity linear combination are stored as a private {\ttfamily double} pointer
of size ${\rm N}^2$. All the right hand side terms of the Von Neumann equation Eq.(\ref{eq:rhoevol_sun_int}) must be 
constructed using {\ttfamily SU\_vector} objects. In order to do this, the {\ttfamily SU\_vector} class implements
operations such as: addition, subtraction, scaling by a constant, rotation, time evolution, and traces. 
Of these the time evolution and rotation are the most computationally expensive. In order to 
improve the efficiency of the code we have implemented algebraic solutions of Eqs.\mref{eq:rotequation,eq:timeequation} that implement rotation and time evolution respectively, and can be found in the files
located in {\ttfamily  SQuIDS/inc/SU\_inc/}.

The headers that declare the class and operatios are in the files {\ttfamily  SQuIDS/inc/SUNalg.h} and
the source code in {\ttfamily  SQuIDS/inc/SUNalg.cpp}. Details of the operation implementation 
can be found in {\ttfamily  SQuIDS/inc/detail/} and {\ttfamily  SQuIDS/inc/SU\_inc/}.

\subsubsection{Constructors}

Different constructors and initialization functions are added in order to 
make more flexible the object initialization.

\begin{itemize}
\item Default constructor.
  \begin{lstlisting}
    SU_vector();
  \end{lstlisting}
  
  Constructs an empty {\ttfamily SU\_vector} with no size.
  
\item Copy constructor.
  \begin{lstlisting}
    SU_vector(const SU_vector& V);
  \end{lstlisting}
  
  The newly constructed {\ttfamily SU\_vector} will allocate its own storage which it will manage automatically.

\item Move constructor.
  \begin{lstlisting}
    SU_vector(SU_vector&& V);
  \end{lstlisting}
  
  If V owned its own storage, it will be taken by the newly constructed {\ttfamily SU\_vector}, and V will be left empty, as if default constructed.

\item Constructor of a zero vector in an N-dimensional Hilbert space .
  
  \begin{lstlisting}
    SU_vector(unsigned int dim);
  \end{lstlisting}
  
  This constructor allocates the memory (which will be managed automatically) for the components of the 
  {\ttfamily SU\_vector} with size given by the argument and initializes it to zero.
  
\item Constructor with a pointer to double.
  
  \begin{lstlisting}
    SU_vector(int dim,double*);
  \end{lstlisting}
  
  This constructor initializes the {\ttfamily SU\_vector} with dimension given by
  the value of {\ttfamily  dim} and uses as the {\ttfamily SU\_vector} components
  the values in the {\ttfamily  double*}.  The newly constructed {\ttfamily SU\_vector} will treat the specified 
  data buffer as its backing storage. The user is responsible for ensuring that this buffer is large
   enough (at least $\mathtt{dim}^2$), and has a lifetime at least as long as the constructed vector and 
   any vectors which inherit the use of this buffer from it by move construction or assignment. 
   The contents of the buffer will not be modified during construction.
  
\item {\ttfamily gsl\_matrix\_complex} constructor
  
  \begin{lstlisting}
    SU_vector(const gsl_matrix_complex*);
  \end{lstlisting}
  Constructs a {\ttfamily SU\_vector} from a complex hermitian {\ttfamily GSL} matrix. The {\ttfamily gsl\_matrix\_complex*}
  contents will copied and not modified.
  
\item Constructor with standard vector.
  
  \begin{lstlisting}
    SU_vector(const std::vector<double>& vector);
  \end{lstlisting}
  
  Constructs {\ttfamily SU\_vector} of dimension equal to the square root of size of the {\ttfamily vector} and components 
  given by the contents of the {\ttfamily vector}. The newly constructed {\ttfamily SU\_vector} will allocate its own storage, 
  but it will copy its component information from {\ttfamily vector}.
  
\end{itemize}

For convenience we also provide factory functions to construct operators that are commonly used such as the identity,
projectors, and generators. The following are available:

\begin{itemize}

\item Identity.
  
  \begin{lstlisting}
    SU_vector Identity(unsigned int dim);
  \end{lstlisting}
  
  Returns a {\ttfamily SU\_vector} which represents the identity of a Hilbert space of dimension {\ttfamily dim}.

\item Generator.
  
  \begin{lstlisting}
    SU_vector Generator(unsigned int dim, unsigned int i);
  \end{lstlisting}
  
  Returns a {\ttfamily SU\_vector} that represents the {\ttfamily i}th generator ($\lambda_i$) of a Hilbert space of dimension {\ttfamily dim}.

\item Projector.
  
  \begin{lstlisting}
    SU_vector Projector(unsigned int dim, unsigned int i);
  \end{lstlisting}
  
  Returns a {\ttfamily SU\_vector} that represents the projector operator in to the subspace 
  spanned by the {\ttfamily i}-state. Namely, $\mathtt{Projector} = {\rm diag}(0,...,0,1,0,...,0)$ where the one is in the {\ttfamily i}th entry.

\item PosProjector.
  
  \begin{lstlisting}
    SU_vector PosProjector(unsigned int dim, unsigned int i);
  \end{lstlisting}
  
  Returns a {\ttfamily SU\_vector} that represents a projector to the upper subspace of dimension {\ttfamily i}, i.e. $\mathtt{PosProjector} =
   {\rm diag}(1,...,1,0,...,0)$ where the last one is in the {\ttfamily i-1} entry.
  
\item NegProjector.
  
  \begin{lstlisting}
    SU_vector NegProjector(unsigned int dim, unsigned int i);
  \end{lstlisting}
  
  Returns a {\ttfamily SU\_vector} that represents a projector to the lower subspace of dimension {\ttfamily i}, i.e. $\mathtt{NegProjector} =
   {\rm diag}(0,...,0,1,...,1)$ where the first one is in the {\ttfamily d-i} entry.
  
\end{itemize}

\subsubsection{Functions}
	
In this section we will describe the general functions that are used to manipulate the objects,
access the values and do different operations.

\begin{itemize}

\item SetAllComponents.

  \begin{lstlisting}
    void SetAllComponents(double v);
  \end{lstlisting}

  Sets all the components of the {\ttfamily  SU\_vector} to {\ttfamily v}.
  
\item SetBackingStore.

  \begin{lstlisting}
    void SetBackingStore(double * storage);
  \end{lstlisting}

  Sets the external storage used by the {\ttfamily SU\_vector}.
  If the {\ttfamily  SU\_vector} had previously allocated automatically managed storage, 
  that memory will be deallocated; but if it had previously used manually specified storage
  it will simply cease using that storage without attempting to deallocate it. All data previously
  stored in the {\ttfamily  SU\_vector} is lost after this function is called. 

\item Rotate function.

  \begin{lstlisting}
    SU_vector& Rotate(unsigned int i,unsigned int j,
                      double theta, double delta) const;
  \end{lstlisting}
  Returns the rotated {\ttfamily  SU\_vector} by a rotation $R_{ij}$
  in the $ij$-subspace by an angle  $\theta_{ij}$ (in radians) and complex phase 
  $\delta_{ij}$ (in radians) defined in Eq.(\ref{eq:mixingdeff}).
  In order to make it efficient and general uses the analytic solution form Eq.(\ref{eq:rotequation}) 
  stored in {\ttfamily  SQuIDS/inc/SU\_inc}. 

\item Change of basis.

  \begin{lstlisting}
    void RotateToB1(const Const& param);
    void RotateToB0(const Const& param);
  \end{lstlisting}

  This functions uses {\ttfamily Rotate} to transform the {\ttfamily SU\_vector} from a basis {\ttfamily B0} to {\ttfamily B1} or viceversa.
  The mixing matrix that defines the unitary transformation that relate {\ttfamily B0} to {\ttfamily B1}, defined in Eqs.\mref{eq:mixingdeff,eq:operator_rot},
  are given by the parameters in {\ttfamily param}. In particular, {\ttfamily RotateToB0} transform a {\ttfamily SU\_vector} from the {\ttfamily B1} basis to the {\ttfamily B0} 
  basis, whereas {\ttfamily RotateToB1} does the opposite.
   
\item Dimension.

  \begin{lstlisting}
    unsigned int Dim(void) const;
  \end{lstlisting}

  Returns the dimension of the Hilbert space.
  
\item Size of vector array.

  \begin{lstlisting}
    unsigned int Size(void) const;
  \end{lstlisting}

  Returns the number of components of the {\ttfamily SU\_vector}.

\item Evolution by time independent hamiltonian.

  \begin{lstlisting}
    SU_vector Evolve(const SU_vector& h0,double t) const;
  \end{lstlisting}

  This function returns a {\ttfamily  SU\_vector} after applying the
  time evolution driven by the operator {\ttfamily h0} argument during a time interval {\ttfamily t}. The analytic solutions of
  Eq.(\ref{eq:timeequation}) stored in {\ttfamily  SQuIDS/inc/SU\_inc/} are used. 
  The {\ttfamily SU\_vector\&} operator must be a diagonal operator, therefore a linear
  combinations of the {\ttfamily  projector} in the {\ttfamily  B0} basis. 
  This choice of the basis makes the evolution more efficient
  since is not necessary to diagonalize the operator every time.
\end{itemize}

\subsubsection{Operators}
Some of the standard C++ operators are overloaded to make more simple and natural writing 
mathematical expressions. Here we list the operators and how they are defined.

\begin{itemize}
\item Logical equality operator ({\ttfamily  ==}).

  \begin{lstlisting}
    bool operator ==(const SU_vector&) const;
  \end{lstlisting}

  Returns {\ttfamily  true} or {\ttfamily  false} if both {\ttfamily  SU\_vectors} are equal.

\item Scalar product operator ({\ttfamily  *}).

  \begin{lstlisting}
    double operator*(const SU_vector&) const;
  \end{lstlisting}

Returns the scalar product of the two vectors, which is equivalent to the trace of the product of the operators that they represent.
 This operation is useful to compute
expectation values, see Eq.\eqref{eq:expected_val}.

\item Product by scalar ({\ttfamily  *}).

  \begin{lstlisting}
    SU_vector operator*(const double) const;
  \end{lstlisting}

  Returns the {\ttfamily  SU\_vector} re-scaled by the {\ttfamily  double}.

\item Assignation operator ({\ttfamily  =})

Assigns the value of the {\ttfamily  SU\_vector} on the right to the one on the left.
The dimensions of the {\ttfamily  SU\_vector} must be the same.

\item Sum operator ({\ttfamily  +}).

  \begin{lstlisting}
    SU_vector operator +(const SU_vector&) const;
  \end{lstlisting}

  Returns the sum of two {\ttfamily  SU\_vector} objects.

\item Subtraction operator ({\ttfamily  -}).
  
  \begin{lstlisting}
    SU_vector operator -(const SU_vector&) const;
  \end{lstlisting}

  Returns the subtraction of two {\ttfamily  SU\_vector} objects.
  
\item Negation operator ({\ttfamily  -}).
  
  \begin{lstlisting}
    SU_vector operator -() const;
  \end{lstlisting}

  Returns the additive inverse of an {\ttfamily  SU\_vector} object.

\item Assignment and move assignment operators ({\ttfamily =})

  \begin{lstlisting}
    SU_vector & operator=(const SU_vector&);
    SU_vector & operator=(SU_vector&&);
  \end{lstlisting}
  
  The assignment operator as well as the move assignment operator are defined. In particular,
  for the move operator, if the vector is empty or owns its own storage it will switch to using whatever
  storage was used by other, causing other to relinquish any ownership of that storage. If, however, 
  the vector is non-empty and uses external storage, it will copy other's data rather than shifting its storage. 
  In this case if the dimensions of the two vectors differ the assignment will fail.

\item Assignment addition and subtraction operators ({\ttfamily  -= +=})

  \begin{lstlisting}
    SU_vector & operator+=(const SU_vector&);
    SU_vector & operator-=(const SU_vector&);
  \end{lstlisting}
  
  These operations combine the addition and subtraction with the
  assignment as usual in C++.
  
\item Assignment multiplication and division operators ({\ttfamily  *= /=})

  \begin{lstlisting}
    SU_vector & operator*=(double);
    SU_vector & operator/=(double);
  \end{lstlisting}
  
  These operations combine the multiplication and division by scalars 
  with the {\ttfamily SU\_vector} assignment operation.

\item Array like component access.

  \begin{lstlisting}
    double& operator[](int);
    const double& operator[](int) const;
  \end{lstlisting}

  Returns the component given by the {\ttfamily  int} argument.

\item Ostream operator ({\ttfamily  <<}).

  \begin{lstlisting}
    friend ostream& operator<<(ostream&, const SU_vector&);
  \end{lstlisting}

  Writes the components of the {\ttfamily  SU\_vector} in to the {\ttfamily  ostream} object as human-readable text.

\end{itemize}

\subsubsection{External functions}

We have also defined quantum mechanical operations between two {\ttfamily SU\_vector}. Furthermore,
to optimize the code we have implemented the {\ttfamily iCommutator} and {\ttfamily Anticommutator}
analytically and stored them in {\ttfamily  SQuIDS/inc/SU\_inc}.

\begin{itemize}
\item iCommutator.
  
  \begin{lstlisting}
    SU_vector iCommutator(const SU_vector&,const SU_vector&);
  \end{lstlisting}
  
  Returns the {\ttfamily  SU\_vector} result of $i$ times the commutator
  of the {\ttfamily  SU\_vector} objects given as an arguments Eq.(\ref{eq:SUN_alg}).

\item Anticommutator.
  
  \begin{lstlisting}
    SU_vector ACommutator(const SU_vector&,const SU_vector&);
  \end{lstlisting}
  
  Returns the {\ttfamily  SU\_vector} result of the anti-commutator
  of the {\ttfamily  SU\_vector}s objects given as an arguments Eq.(\ref{eq:SUN_alg-2}).
  
\item Trace function.
  \begin{lstlisting}
    double SUTrace(const SU_vector&, const SU_vector&);
  \end{lstlisting}

  Returns the trace of the product of the two operators represented by the {\ttfamily  SU\_vectors} given
  in the arguments. It is the same as the scalar product.
\end{itemize}

\subsubsection{Usage and optimization}

A {\ttfamily SU\_vector} of dimension $N$ has $N^2$ components that are stored as a private {\ttfamily double*}. By default {\ttfamily SU\_vector} will automatically allocate sufficient `backing storage' to contain these. It is, however, possible to specify that an {\ttfamily SU\_vector} should treat some externally provided buffer as its backing storage. In this case the size and lifetime of that buffer are the responsibility of the user, so users are encouraged to avoid using this mode unless it is required by their application, as its use is more difficult and requires much greater care. The external storage mode is primarily useful because it allows interfacing with other, low-level numerical codes efficiently.

As described in the previous sections {\ttfamily SU\_vector} provides overloaded mathematical operators so that algebra can be written in a natural way. Furthermore, the SQuIDS library has a limited ability to optimize away temporary objects (via a partial expression template system). That is, all operations of the forms 

\begin{lstlisting}
v1 [Op1]= v2 [Op2] v3;
v1 [Op1]= s * v2;
\end{lstlisting}
where {\ttfamily v1}, {\ttfamily v2}, and {\ttfamily v3} are pre-existing objects of type {\ttfamily SU\_vector} (and {\ttfamily s} is a scalar) are performed without allocating memory. {\ttfamily Op1} may {\ttfamily+},{\ttfamily -}, or nothing (normal assignment), and {\ttfamily Op2} may be {\ttfamily +}, {\ttfamily -}, time evolution, a commutator or an anticommutator.

This optimization is inhibited when {\ttfamily v1} aliases {\ttfamily v2} or {\ttfamily v3} (they are the same objects or they otherwise refer to the same backing storage) and the operation being performed involves components in the input and output vectors with different indices. This has no influence on the correctness of writing complex expressions in terms of subexpressions: These will still be correctly evaluated, but memory will be allocated for the results of the subexpressions, making the calculation slower than if this can be avoided. It is expected that users will write expressions in the form they find most natural, and only if performance optimization is required consider restructuring code to take deliberate advantage of this optimization. In that case, the following techniques may be useful:

\begin{itemize}
\item If a calculation involving subexpressions is performed in a loop, it is advantageous to manually create sufficient temporaries for all subexpressions outside of the loop and then split the complex expression into a series of basic operations whose results are stored to the temporaries. This ensures that allocation will be performed only once per temporary before entering the loop, rather than once per temporary, per loop iteration. For example:
\begin{lstlisting}[frame=leftline, numbers = left,breaklines=true, label = suv:opt11]
//assuming size N arrays of SU_vector state, v1, v2, v3, and v4
//and a floating-point t
for(unsigned int i=0; i<N; i++)
    state[i] += v1[i].Evolve(v2[i],t) 
              + v3[i].Evolve(v4[i],t);
\end{lstlisting}

In this code the addition on the right hand side can be performed without allocation, but each of the evolution operations must allocate a temporary, so 2*N allocations and deallocations must occur. This can be reduced to 2 allocations and deallocations by rewriting in this form:

\begin{lstlisting}[frame=leftline, numbers = left,breaklines=true, label = suv:opt12]
SU_vector temp1, temp2;
for(unsigned int i=0; i<N; i++){
  temp1 = v1[i].Evolve(v2[i],t);
  temp2 = v3[i].Evolve(v4[i],t)
  state[i] += temp1 + temp2;
}
\end{lstlisting}

\item If a calculation has an {\ttfamily SU\_vector} calculation as a subexpression, but otherwise operates on scalars, it can be useful to rewrite the expression so that the vector calculation forms the top level if possible:

\begin{lstlisting}[frame=leftline, numbers = left,breaklines=true, label = suv:opt2]
//assuming SU_vectors v1 and v2 and scalars s1 and s2
v1 = s1*(s2*v2);
//can be better reassociated as:
v1 = (s1*s2)*v2;
\end{lstlisting}

\end{itemize}

\subsection{SQUIDS}

This object implements the numerical solution for a set of density matrices plus a set 
of scalar functions in the interaction picture, where the evolution given by $H_0$ is solved analytically.
The numerical calculation is done using the GNU Scientific Library, and different parameters 
for the numerical precision and integrator can be set through the SQuIDS interface.

\begin{figure}[h]
\centering
\includegraphics{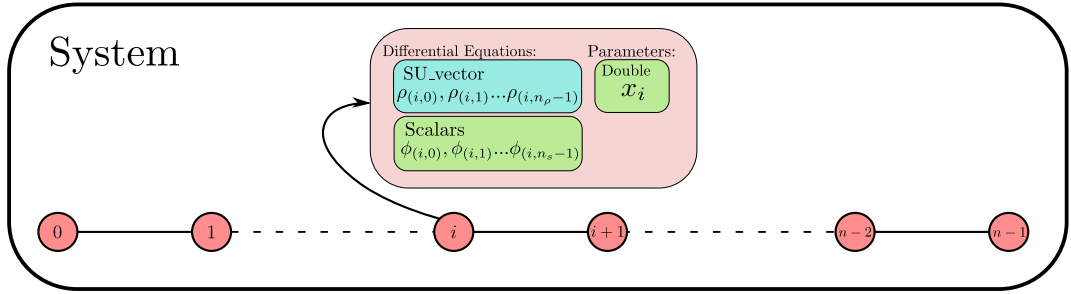}
\caption{Scheme of the distribution of density matrix and scalar
  functions on the nodes}
\label{fig:squids}
\end{figure}

As we described before in Sec. \ref{sec:system_time_def} the system consist of a set of nodes 
where every $i$-node contains $n_\rho$ density matrix expressed as {\ttfamily SU\_vector}, $n_s$ scalar functions which are 
{\ttfamily double}, and a {\ttfamily double} parameter $x_i$. The scheme is shown in Fig.\ref{fig:squids}.

The object is defined so that all of the terms in the right hand side of the differential 
equation are defined as virtual functions that by default return zero, and can be overridden by the user.

If the user does not activate the terms that are numerically solved, then the evolution will be done analytically.

\subsubsection{Constructors and Initializing Functions}

\begin{itemize}

\item Default constructor.
    
  \begin{lstlisting}
    SQUIDS();
  \end{lstlisting}
    
Constructs an uninitialized {\ttfamily  SQUIDS} object.

\item Constructor and initializing function.

  \begin{lstlisting}
    SQUIDS(unsigned int nx,unsigned int dim,unsigned int nrho,
           unsigned int nscalar, double ti = 0);
    void ini(unsigned int nx,unsigned int dim,unsigned int nrho,
           unsigned int nscalar, double ti = 0);
  \end{lstlisting}
 
  Initializes the object, allocating all necessary memory for the density matrices and scalars.
  The arguments are described in Table \ref{tbl:squids_init}.
  
  \begin{table}[h]
    \begin{tabular}{|c|c|p{12cm}|}
      \hline
      Argument & Type & description \\ \hline \hline
      {\ttfamily nx}  & {\ttfamily unsigned int} &  Number of nodes in the problem, each node
      has a set of density matrices and scalars as defined by the other arguments. \\ \hline
      {\ttfamily dim}  & {\ttfamily unsigned int} &Dimension of the Hilbert space of the density matrices. \\ \hline
      {\ttfamily nrho}  & {\ttfamily unsigned int} &Number of density matrices in each node. \\ \hline
      {\ttfamily nscalar}  & {\ttfamily unsigned int} &Number of scalars functions in each node. \\ \hline
      {\ttfamily ti}  & {\ttfamily double} & Initial time of the system. (Defaults to zero.) \\ \hline
    \end{tabular}
    \caption{Arguments description of SQUIDS constructor and
      initialization function from left to right.}
    \label{tbl:squids_init}
  \end{table}
    
\end{itemize}

\subsubsection{Functions}

\begin{itemize}
\item Set the range for the array {\ttfamily  x}.

  \begin{lstlisting}
    int Set_xrange(double xini, double xend, string scale);
  \end{lstlisting}

  This function sets the values on the array {\ttfamily  x}. The array will start
  at {\ttfamily xini} and end at {\ttfamily xend} (inclusive) with either a uniform linear or 
  logarithmic spacing.

  \begin{table}[h]
    \begin{tabular}{|c|c|p{11cm}|}
      \hline
      Argument & type & description \\ \hline \hline
      {\ttfamily xini}  & {\ttfamily double} & smaller value of {\ttfamily  x}. \\ \hline
      {\ttfamily xend}  & {\ttfamily double} & largest value of  {\ttfamily  x}. \\ \hline
      {\ttfamily scale}  & {\ttfamily string} & Either {\ttfamily  "lin"} or {\ttfamily  "log"} and sets the scale 
      as a linear or logarithmic. \\ \hline
    \end{tabular}
    \caption{Arguments of the {\ttfamily Set\_xrange} function.}
  \end{table}

\item Get value of {\ttfamily  x}

  \begin{lstlisting}
    double Get_x(unsigned int i) const;
  \end{lstlisting}
  Returns the {\ttfamily i}th value of the {\ttfamily  x} array.

\item Get the bin in {\ttfamily  a}

  \begin{lstlisting}
    int Get_i(double x) const;
  \end{lstlisting}

  Returns the index of the {\ttfamily  x} array whose value is closest to {\ttfamily a}. 
  {\ttfamily  a} must be between the values previously specified for {\ttfamily xini} and {\ttfamily xend} 
  in the most recent call to {\ttfamily Set\_xrange}. 
  
\item Get current system time

  \begin{lstlisting}
    double Get_t() const;
  \end{lstlisting}
  Returns the current time of the system. 
  
\item Get {\ttfamily params} object

  \begin{lstlisting}
    double Get_t_initial() const;
  \end{lstlisting}
  Returns the initial time of the system.
  
\item Get initial time

  \begin{lstlisting}
    const * Const Get_params() const;
  \end{lstlisting}
  Returns a const reference to the {\ttfamily params} SQUIDS protected member.
  
\item Derivative.

  \begin{lstlisting}
    int Derive(double t);
  \end{lstlisting}

  Computes the derivative, r.h.s. of Eq.\mref{eq:rhoevol_sun_int,eq:rhoevol_sun_int-2}, of the system at a time {\ttfamily t}, including all of the terms defined by the user by overriding the virtual functions specified in section \ref{ssec:squids_virt_func}. Note that for each user-supplied interaction term the appropriate flag must be set {\ttfamily true} for that term to be included (see Sec. \ref{sec:set_function} and Table \ref{tbl:squids_set_virtual}). 

\item Evolution function.

  \begin{lstlisting}
    int Evolve(double dt);
  \end{lstlisting}

  Evolves the system by a time interval {\ttfamily dt} given in natural units \footnote{We set $c = \hbar = k_b = 1$.}  (see {\ttfamily Const} class in Sec. \ref{sec:const} and, in particular, Table \ref{tbl:const_prop}).

\item Get expectation value.

  \begin{lstlisting}
    double GetExpectationValue(SU_vector op,unsigned int irho,
                               unsigned int ix) const;
    double GetExpectationValueD(SU_vector op,unsigned int irho,
                               double x) const;
  \end{lstlisting}

  The first function returns the expectation value of the operator represented by {\ttfamily op} for the state
  in the {\ttfamily irho} density matrix in that {\ttfamily ix} node at the current time {\ttfamily t}. Notice
  that {\ttfamily op} is evolved by the $H_{0(ix,irho)}$ hamiltonian in order to go the interaction picture at time {\ttfamily t}.
  
  In general $H_{0J}$ may depend continuously in the parameter $x$, in that case in order to compute the expectation value it is very useful 
  perform the evolution of the operator {\ttfamily op} driven by $H_{0J}$ for the exact value of $x$. The second function uses this method
  together with a linear interpolation in $\rho$ over the parameter $x$ in order to give the expectation value.
   
\end{itemize}

\subsubsection{Virtual Functions}\label{ssec:squids_virt_func}

All of these virtual functions return a zero {\ttfamily  SU\_vector} by default and may be overridden by the user in a derived class to define a problem. When implemented, each function's return value must be in natural units (see Table \ref{tbl:const_prop}).

\begin{itemize}
\item Time independent Hamiltonian $H_{0J}$.

  \begin{lstlisting}
    virtual SU_vector H0(double x,unsigned int irho) const;
  \end{lstlisting}

  This function returns the time independent hamiltonian that will be solved 
  analytically, for a particular value of the parameter {\ttfamily x} and density matrix {\ttfamily irho}. 
  It is important to note that the analytic solution is implemented
  assuming that this operator is represented by a diagonal matrix; therefore the problem basis
  should be chosen to satisfy this condition. For example, in the case of neutrino oscillations
  this means that the operators are defined in the mass basis.

  The {\ttfamily  double x} gives the parameter corresponding to the node of the {\ttfamily  x} array. Notice that $H_{0J}$ 
  must be defined as a continuous function of {\ttfamily x} since $H_{0J}$ is solved analytically and thus its independent 
  of the discrete nodes, as this allows {\ttfamily GetExpectationValueD} to calculate observables at arbitrary {\ttfamily x}.

\item Time dependent Hamiltonian $H_{1J}$.

  \begin{lstlisting}
   virtual SU_vector HI(unsigned int ix,unsigned int irho,
                        double t) const;
  \end{lstlisting}

  Returns the time dependent part of the Hamiltonian
  at the {\ttfamily ix} node for the density matrix {\ttfamily irho} at time {\ttfamily  t}. The result of this function is used to calculate the commutator 
  that drives the quantum evolution, the first term in Eq.\eqref{eq:rhoevol_sun_int}. 
  %TODO: is the above equation reference correct?
  
  This function is used only if {\ttfamily Set\_CoherentRhoTerms} has been called with a {\ttfamily true} argument. 

\item Non coherent terms $\Gamma_J$.

  \begin{lstlisting}
    virtual SU_vector GammaRho(unsigned int ix,
                           unsigned int irho, double t) const;
  \end{lstlisting}

  Returns the non-coherent interaction at the {\ttfamily ix} node for the density matrix {\ttfamily irho} at time {\ttfamily  t}. 
  The result of this function is used
  to calculate the anticommutator, the second term in Eq.\eqref{eq:rhoevol_sun_int}.
  
  This function is used only if {\ttfamily Set\_NonCoherentRhoTerms} has been called with a {\ttfamily true} argument. 
   
\item Other $\rho$ interactions $F_J$.

  \begin{lstlisting}
    virtual SU_vector InteractionsRho(unsigned int ix,
                           unsigned int irho,double t) const;
  \end{lstlisting}

  Returns the third term on the right hand side of
  Eq.(\ref{eq:rhoevol_sun_int}) at the {\ttfamily ix} node for the density matrix {\ttfamily irho} at time {\ttfamily  t}.
  For example, terms mixing different density matrices and scalar functions of different 
  nodes can be included here.
  
  This function is used only if {\ttfamily Set\_OtherRhoTerms} has been called with a {\ttfamily true} argument. 
  
\item Scalar function attenuation $\Gamma_K$.

  \begin{lstlisting}
    virtual double GammaScalar(unsigned int ix,
                         unsigned int iscalar, double t) const;
  \end{lstlisting}
  
  It returns the Boltzmann attenuation term (first term) for the scalar function
  Eq.(\ref{eq:rhoevol_sun_int}) at the {\ttfamily ix} node for the scalar function {\ttfamily iscalar} at time {\ttfamily  t}.
  
  This function is used only if {\ttfamily Set\_GammaScalarTerms} has been called with a {\ttfamily true} argument. 

\item Other scalar interactions $G_K$.

  \begin{lstlisting}
    virtual double InteractionsScalar(unsigned int ix,
                         unsigned int iscalar,double t) const;
  \end{lstlisting}
  
  Returns any necessary second term on the right hand side of
  Eq.(\ref{eq:rhoevol_sun_int}) at the {\ttfamily ix} node for the scalar function {\ttfamily iscalar} at time {\ttfamily  t}.
  This may include terms that depend on the other scalars and density matrices.
  
  This function is used only if {\ttfamily Set\_OtherScalarTerms} has been called with a {\ttfamily true} argument. 

\item Pre-Derivative Function.
  
  \begin{lstlisting}
    virtual void PreDerive(double t);
  \end{lstlisting}

  This function is called every time before computing the derivatives at time {\ttfamily t}, i.e. before evaluating 
  the virtual functions described above, and can be used to pre-calculate variables that will be used in the derivative
  or time evolve projectors in order to more easily and efficiently compute the preceding functions. In general,
  any update of the parameters in the preceding functions can be included.
  For example, in the case of neutrino oscillations, this can include the evolution of the flavor projectors that can be 
  then used to define the $H_{1J}$ Hamiltonian.

  \end{itemize}

  \begin{table}[h]
  \begin{center}
    \begin{tabular}{|c|p{8cm}|}
      \hline
      Set Function & Description \\ \hline \hline
      {\ttfamily Set\_CoherentRhoTerms} &  activate the use of  {\ttfamily HI} \\ \hline
      {\ttfamily Set\_NonCoherentRhoTerms}  & activate the use of  {\ttfamily GammaRho} \\ \hline
      {\ttfamily Set\_OtherRhoTerms}  & activate the use of  {\ttfamily InteractionsRho} \\ \hline
      {\ttfamily Set\_GammaScalarTerms}  & activate the use of  {\ttfamily GammaScalar} \\ \hline
      {\ttfamily Set\_OtherScalarTerms}  & activate the use of  {\ttfamily InteractionsScalar} \\ \hline
    \end{tabular}
    \caption{Set functions that control the use of virtual functions.}
    \label{tbl:squids_set_virtual}
    \end{center}
  \end{table}

\subsubsection{Set functions}
\label{sec:set_function}

The set functions configure different parameters in the {\ttfamily  SQUIDS} object.

\begin{itemize}
\item Set {\ttfamily GSL} stepper function.

  \begin{lstlisting}
    void Set_GSL_set(const gsl_odeiv2_step_type * opt);
  \end{lstlisting}

Sets the {\ttfamily GSL} stepper function for the differential numerical algorithm, see Table (\ref{tbl:gsl-stepper}). Note that this is a 
subset of the methods supported by {\ttfamily GSL} and only includes those that do not require second derivative. By default
{\ttfamily gsl\_odeiv2\_step\_rkf45} is set. For more details see the {\ttfamily GSL} website \cite{GSL}.

\begin{center}
\begin{table}[h]
\begin{center}
    \begin{tabular}{|c|}
    \hline \hline
    Available {\ttfamily GSL} stepper functions \\ \hline \hline
          {\ttfamily gsl\_odeiv2\_step\_rk2} \\ \hline
           {\ttfamily gsl\_odeiv2\_step\_rk4} \\ \hline
           {\ttfamily gsl\_odeiv2\_step\_rkf45} \\ \hline
      {\ttfamily gsl\_odeiv2\_step\_rkck} \\ \hline
      {\ttfamily gsl\_odeiv2\_step\_rk8pd} \\ \hline
      {\ttfamily gsl\_odeiv2\_step\_msadams} \\ \hline
  \end{tabular}
\end{center}
  \caption{Available {\ttfamily GSL} stepper functions.}
  \label{tbl:gsl-stepper}
\end{table}
\end{center}

\item Set absolute error

  \begin{lstlisting}
    void Set_abs_error(double error);
  \end{lstlisting}
  
  Sets the {\ttfamily GSL} algorithm absolute error to {\ttfamily error}.

\item Set relative error

  \begin{lstlisting}
    void Set_rel_error(double error);
  \end{lstlisting}
  
   Sets the {\ttfamily GSL} algorithm relative error to {\ttfamily error}.

\item Set initial step

  \begin{lstlisting}
    void Set_h(double h);
  \end{lstlisting}
  
   Sets the {\ttfamily GSL} algorithm initial step to {\ttfamily h}.

\item Set minimum step

  \begin{lstlisting}
    void Set_h_min(double h);
  \end{lstlisting}
  
   Sets the {\ttfamily GSL} algorithm minimum step to {\ttfamily h}.

\item Set maximum step

  \begin{lstlisting}
    void Set_h_max(double h);
  \end{lstlisting}
  
   Sets the {\ttfamily GSL} algorithm maximum step to {\ttfamily h}.

\item Switch adaptive stepping

  \begin{lstlisting}
    void Set_AdaptiveStep(bool opt);
  \end{lstlisting}
  
  If {\ttfamily opt} is {\ttfamily true} adaptive stepping will be used, otherwise a fixed step size will be used. In the fixed
  step case the number of steps can be set by  {\ttfamily Set\_NumSteps}.
  
\item Set number of steps (for fixed stepping)

  \begin{lstlisting}
    void Set_NumSteps(int steps);
  \end{lstlisting}
  
 Sets the number of steps used when not using adaptive stepping.

\item Switch Coherent Interactions

  \begin{lstlisting}
    void Set_CoherentRhoTerms(bool opt);
  \end{lstlisting}
  
  If {\ttfamily opt} is {\ttfamily true} the implemented {\ttfamily HI} will be used, otherwise it will be ignored and treated as zero.

\item Switch Non Coherent Interactions

  \begin{lstlisting}
    void Set_NonCoherentRhoTerms(bool opt);
  \end{lstlisting}
  
  If {\ttfamily opt} is {\ttfamily true} the implemented {\ttfamily GammaRho} will be used, otherwise it will be ignored and treated as zero.

\item Switch extra $\rho$ Interactions

  \begin{lstlisting}
    void Set_OtherRhoTerms(bool opt);
  \end{lstlisting}
  
  If {\ttfamily opt} is {\ttfamily true} the implemented {\ttfamily InteractionsRho} will be used, otherwise it will be ignored and treated as zero.

\item Switch Scalar Attenuation.

  \begin{lstlisting}
    void Set_GammaScalarTerms(bool opt);
  \end{lstlisting}
  
  If {\ttfamily opt} is {\ttfamily true} the implemented {\ttfamily GammaScalar} will be used, otherwise it will be ignored and treated as zero.
  
\item Switch Scalar Interactions

  \begin{lstlisting}
    void Set_OtherScalarTerms(bool opt);
  \end{lstlisting}
  
  If {\ttfamily opt} is {\ttfamily true} the implemented {\ttfamily InteractionsScalar} will be used, otherwise it will be ignored and treated as zero.

\end{itemize}

\subsubsection{Protected members}
\label{sec:params}

In this section we describe some useful protected member of the SQUIDS derive which derive
classes may want to use.

\begin{itemize}

\item {\ttfamily Const} object

  \begin{lstlisting}
    Const params;
  \end{lstlisting}
  
  {\ttfamily Const} object that contains useful units and angles that define the transformation between basis.

\item State of a node

  \begin{lstlisting}
    struct SU_state {
    	std::unique_ptr<SU_vector[]> rho;
	double* scalar;
    };
  \end{lstlisting}
  
  Contains the state of a node consisting of density matrices contained in {\ttfamily rho} and scalar functions in {\ttfamily scalar}. 

\item System state

  \begin{lstlisting}
    std::unique_ptr<SU_state[]> state;
  \end{lstlisting}
  
  Contains the state of the system at the current time, namely the density matrices and scalars at all nodes.
  
\item Number of nodes

  \begin{lstlisting}
    unsigned int nx;
  \end{lstlisting}
  
  Number of nodes in the system.
  
\item Number of scalars

  \begin{lstlisting}
    unsigned int nscalars;
  \end{lstlisting}
  
  Number of scalar functions per node.
  
\item Number of density matrices.

  \begin{lstlisting}
    unsigned int nrhos;
  \end{lstlisting}
  
  Number of density matrices per node.
  
\item Hilbert space dimension

  \begin{lstlisting}
    unsigned int nsun;
  \end{lstlisting}
  
  Dimension of the Hilbert space of the density matrices. 

\end{itemize}

\section{Included examples}
\label{Examples}
\subsection{Vacuum neutrino oscillations}
\label{sec:vacuum}
This is a basic example in which vacuum neutrino oscillations  
are implemented. In this case, the code does not use any numerical integration,
just the analytic solutions given by the {\ttfamily  SU\_vector} class.

The evolution is defined by $H_0$, which in the mass basis has the following form

\begin{align}
  H_0 = \frac{1}{2E}\begin{pmatrix}
   0 & 0  & 0 \\
   0 & \Delta m^2_{21}  & 0  \\
   0  & 0 & \Delta m^2_{31}  \\
  \end{pmatrix}, 
  \label{eq:dm2}
\end{align}

where the mixing matrix Eq.\eqref{eq:mixing}, which relates the mass and flavor bases,
depends on three mixing angles and
one complex phase: $\{\theta_{12}, \theta_{23}, \theta_{13}, \delta \}$.

In the following, we describe the derived class and the functions
that implement this example. 

\subsubsection{Derived Class ({\ttfamily vacuum})}

This class is defined in  {\ttfamily  SQuIDS/examples/VacuumNeutrinoOscillations/vacuum.h } and
implemented in {\ttfamily  SQuIDS/examples/VacuumNeutrinoOscillations/vacuum.cpp}.

The {\ttfamily vacuum} class constructor has the following signature 

\begin{lstlisting}
    vacuum(unsigned int nbins,unsigned int nflavor,
           double Eini, double Efin);
\end{lstlisting}

where {\ttfamily nbins} is the number of energy bins in a logarithm scale from a minimum {\ttfamily Eini} to a maximum {\ttfamily Efin} 
and {\ttfamily nflavor} is the number of neutrino flavors. On the constructor it calls the SQUIDS {\ttfamily ini} function  in the following way

\begin{lstlisting}
    ini(nbins,nflavor,1,0,0.);
\end{lstlisting}

which initializes {\ttfamily nbins} $x$ nodes, which refer to the neutrino energy, with one
 {\ttfamily SU\_vector} of dimension {\ttfamily nflavor} and no scalar functions. The final parameter
 sets the initial time of the system to zero. Furthermore, 
in the constructor, it is useful to define the projectors in the
flavor and mass basis given by Eq.\eqref{eq:proj}, since $H_0$ is a
linear combination of the mass projectors and the flavor projectors
are needed to evaluate flavor expectation values. The projectors are stored in the following arrays

\begin{lstlisting}
  std::unique_ptr<SU_vector[]> b0_proj;
  std::unique_ptr<SU_vector[]> b1_proj;
\end{lstlisting}
where the {\ttfamily b0} label corresponds to the mass basis
and the {\ttfamily b1} to the flavor basis. In order to define the
transformation between the mass and flavor basis we use {\ttfamily params.Set\_MixingAngle}.

\begin{lstlisting}
  params.SetMixingAngle(0,1,33.48*params.degree); //theta 1,2 
  params.SetMixingAngle(0,2,8.55*params.degree); //theta 1,3 
  params.SetMixingAngle(1,2,42.3*params.degree); //theta 2,3 
\end{lstlisting}
Next we construct the {\ttfamily SU\_vector DM2} that represent the matrix 
in equation Eq.\eqref{eq:dm2}. 

\begin{lstlisting}
  const double ev2=params.eV*params.eV;
  params.SetEnergyDifference(1,7.5e-5*ev2); //delta m^2 2,1
  params.SetEnergyDifference(2,2.45e-3*ev2); //delta m^2 3,1
  
  for(int i = 1; i < nsun; i++)
  	DM2 += (b0_proj[i])*params.GetEnergyDifference(i);
\end{lstlisting}
Finally, we set the initial state of system to the first flavor ($\nu_e$) by means
of the flavor projectors

\begin{lstlisting}
  for(int ei = 0; ei < nx; ei++)
    state[ei].rho[0]=b1_proj[0];
\end{lstlisting}

The next member function is

\begin{lstlisting}
  SU_vector H0(double E, unsigned int irho) const;
\end{lstlisting}
which returns the value of the time independent
Hamiltonian $H_0$. The last member function is defined to get the flux of a given flavor

\begin{lstlisting}
  double Get_flux(unsigned int flavor,double enu);
\end{lstlisting}
where {\ttfamily flavor} specifies the neutrino
flavor and {\ttfamily enu} the neutrino energy.

\subsubsection{Main file}

\begin{figure}[h]
\centering
  \includegraphics{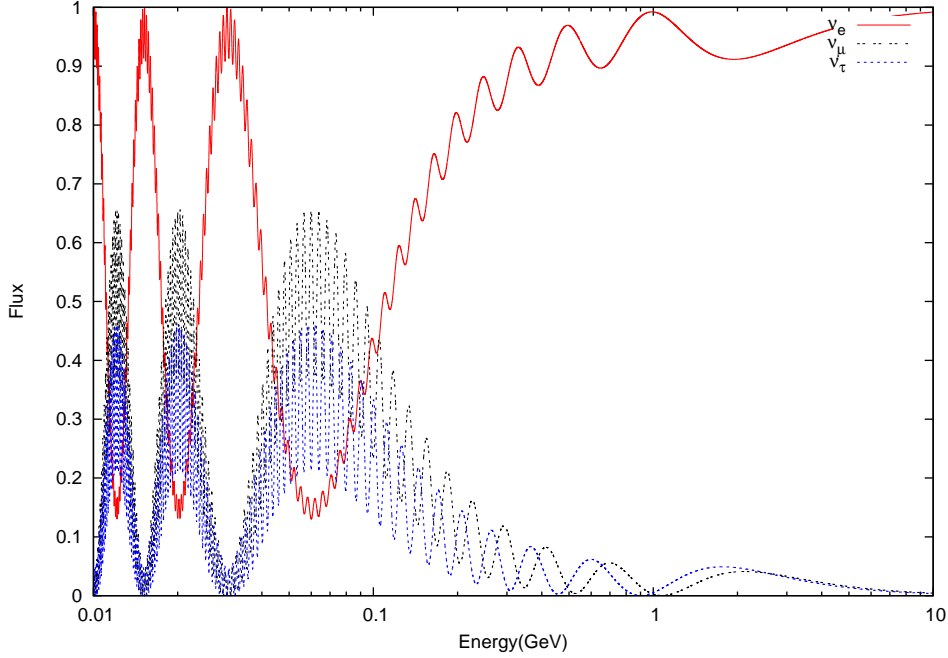}
  \caption{Probability for a neutrino to interact as a particular flavor as a function of energy after propagating 1000 km, starting from a pure $\nu_e$ flux.}
\label{Plotvacuum}
\end{figure}

The main file declares the object and propagates the three standard
neutrino states in a $1000{\rm km}$ baseline. The final flavor
content is saved in the
file {\ttfamily  oscillations.dat}. The output is shown in Fig.\ref{Plotvacuum}.

\begin{lstlisting}[frame=leftline, numbers = left,breaklines=true]
int main(){
  Const units;
  
  //Number of energy bins
  unsigned int Nenergy=1000;
  //Number of flavors
  unsigned int Nflavor=3;
  //Energy Range
  double Emin=10*units.MeV, Emax=10*units.GeV;
  //declaration of the object
  vacuum V0(Nenergy,Nflavor,Emin,Emax);

  V0.Evolve(1000*units.km);

  std::ofstream file("oscillations.dat");

  const int nu_e=0, nu_mu=1, nu_tau=2;
  for(double lE=log(Emin); lE<log(Emax); lE+=0.0001){
    double E=exp(lE);
    file << E/units.GeV << "  " << V0.Get_flux(nu_e,E) << "  " <<
      V0.Get_flux(nu_mu,E) << "  " << V0.Get_flux(nu_tau,E) << std::endl;
  }

  std::cout << std::endl <<  "Done! " << std::endl <<  "Do you want to run the gnuplot script? yes/no" << std::endl;
  std::string plt;
  std::cin >> plt;

  if(plt=="yes" || plt=="y"){
    return system("./plot.plt");
  }
  
  return 0;
}
\end{lstlisting}

\subsection{Rabi oscillations}
\label{sec:Rabi}
This example illustrates the numerical time dependent Hamiltonian solution.
The code solves for the population of a two level system as a function of time under the
influence of an external oscillating potential (e.g. a laser).

We consider two cases: in the first the frequency of the laser is
resonant with the energy difference of the two level system, and in the
second the laser has a small de-tuning.

The evolution of the Rabi system is
driven by the Hamiltonian

\begin{align}
  H(t) = H_0+H_1(t) = 
  \begin{pmatrix}
    \epsilon_1 & 0  \\
    0 & \epsilon_2  \\
  \end{pmatrix} 
  +
  \begin{pmatrix}
    0 & d  \\
    d  & 0  \\
  \end{pmatrix} \xi(t) ,   
\end{align}
where $\epsilon_i$ is the energy of the $i$-state, $d$ is the dipole
expected value, and the function $\xi(t)$ plays the role
of an external laser acting on the system, which is given by

\begin{equation}
\xi(t)= A \cos(\omega t).
\end{equation}

In terms of the $H_0$ eigenstates projectors, $\Pi_i$, the full
Hamiltonian has the form

\begin{equation}
H(t) = \Pi_2 (\epsilon_2 -\epsilon_1) + (U^\dagger(\pi/4) \Pi_1
U(\pi/4) - U^\dagger(\pi/4) \Pi_2 U(\pi/4)) \xi(t), 
\label{eq:rabihamiltonian}
\end{equation}
where the dipole operator is constructed by a linear combination of
rotated projectors.

\subsubsection{Derived class ({\ttfamily rabi})}

The class is declared in {\ttfamily  SQuIDS/examples/RabiOscillations/rabi.h} 
and implemented in {\ttfamily  SQuIDS/examples/RabiOscillations/rabi.cpp}.

As in section \ref{sec:vacuum} the projectors are included, where {\ttfamily
  b0} and {\ttfamily b1} label the projectors to the $H_0$ and dipole eigenstates respectively.
\begin{lstlisting}
  std::unique_ptr<SU_vector[]> b0_proj;
  std::unique_ptr<SU_vector[]> b1_proj;
\end{lstlisting}

It is also useful to define

\begin{lstlisting}
  SU_vector suH0;
  SU_vector d0;
  SU_vector d;
\end{lstlisting}
where {\ttfamily d0} is the dipole operator, {\ttfamily d} is the time evolved dipole operator, and
 {\ttfamily suH0} is the time independent Hamiltonian.

The initialization and constructor depend on the energy
difference of the two levels, {\ttfamily D\_E}, the frequency of the external potential, {\ttfamily w\_i}, 
and the amplitude of the external interaction, {\ttfamily Am}.

\begin{lstlisting}
  rabi(double D_E, double wi, double Am);
  void init(double D_E, double wi, double Am);
\end{lstlisting}

First we initialize the base SQUIDS object for this problem
\begin{lstlisting}
  ini(1/*nodes*/,2/*SU(2)*/,1/*density matrices*/,0/*scalars*/);
\end{lstlisting}
Next, we set the physical parameters of the problem: We use
{\ttfamily SetEnergyDifference} to set the levels' energy difference and
{\ttfamily SetMixingAngle} to construct the transformation between the two
bases, which is $U$ in Eq.\eqref{eq:rabihamiltonian}.
\begin{lstlisting}
  params.SetEnergyDifference(1,D_E);
  params.SetMixingAngle(0,1,params.pi/4);
\end{lstlisting}
We turn on the treatment of coherent terms ({\ttfamily HI}):
\begin{lstlisting}
  Set_CoherentRhoTerms(true);
\end{lstlisting}
We construct the time independent hamiltonian using energy difference, the first term in Eq.\eqref{eq:rabihamiltonian}:
\begin{lstlisting}
  suH0 = b0_proj[1]*params.GetEnergyDifference(1);
\end{lstlisting}
We construct the dipole operators: {\ttfamily d0} is the initial dipole operator and {\ttfamily d} the time evolved
one, which are defined initially to be
\begin{lstlisting}
  d0=(b1_proj[0]-b1_proj[1]);
  d=d0;
\end{lstlisting}
Finally, the system is initialized to the ground state.
 
We also defined the {\ttfamily PreDerive} function

\begin{lstlisting}
  void rabi::PreDerive(double t){
    d=d0.Evolve(suH0,t-Get_t_initial());
  }
\end{lstlisting}
 which evolves the dipole operator to the necessary time. In this problem, {\ttfamily H0} is just
 defined to return the constant {\ttfamily SU\_vector suH0} which was computed at initialization. 
Finally, {\ttfamily HI} is returns the second term in equation Eq.\eqref{eq:rabihamiltonian}
\begin{lstlisting}
  SU_vector rabi::HI(unsigned int ix,
                     unsigned int irho, double t) const{
    return (A*cos(w*t))*d;
  }
\end{lstlisting} 

\subsubsection{Main file}

The main file declares two {\ttfamily  rabi} objects {\ttfamily R0}
and {\ttfamily R1}, the first with two matching frequencies and the second 
with with two frequencies de-tuned by the value given by the user.

\begin{lstlisting}[frame=leftline, numbers = left,breaklines=true]
int main(){
  // Declaration of the objects
  rabi R0,Rd;
  // de-tuning
  double del;

  // delta time for the prints
  double dt=0.01;
  // Final time
  double tf=120;

  // Tuned Rabi system
  R0.init(10,10,0.1);

  // Setting the errors
  R0.Set_rel_error(1e-5);
  R0.Set_abs_error(1e-5);

  std::cout << "Rabi system with frequency of 10 initialized." << std::endl;
  std::cout << "give the value for the detuning: " << std::endl;
  std::cin >> del;

  // un-tuned Rabi system
  Rd.init(10,10+del,0.1);
  // Setting the errors
  Rd.Set_rel_error(1e-5);
  Rd.Set_abs_error(1e-5);

  std::cout << "Computing rabi" << std::endl;
  std::ofstream file("rabi.dat");

  // Evolve and save the evolution
  for(double t=0;t<tf;t+=dt){
    progressbar(100*t/tf);
    R0.Evolve(dt);
    file << t << "\t" << R0.GetExpectationValue(R0.d0,0,0) << "  " 
	 << R0.GetExpectationValue(R0.b0_proj[0],0,0) << "  "
	 << R0.GetExpectationValue(R0.b0_proj[1],0,0) << std::endl;
  }
  file.close();
  file.open("rabi_detuned.dat");
  std::cout << std::endl << "Computing detuned rabi" << std::endl;
  for(double t=0;t<tf;t+=dt){
    progressbar(100*t/tf);
    Rd.Evolve(dt);
    file << t << "\t" << Rd.GetExpectationValue(Rd.d0,0,0) << "  " 
	 << Rd.GetExpectationValue(Rd.b0_proj[0],0,0) << "  " 
	 << Rd.GetExpectationValue(Rd.b0_proj[1],0,0) << std::endl;
  }
  file.close();
  
  //Ask whether to run the gnuplot script
  std::string plt;
  std::cout << std::endl <<  "Done! " << std::endl <<
   "  Do you want to run the gnuplot script? yes/no" << std::endl;
  std::cin >> plt;
  if(plt=="yes" || plt=="y")
    return system("./plot.plt");
  return 0;
}
\end{lstlisting}

\begin{figure}[h!]
 \centering
    \includegraphics{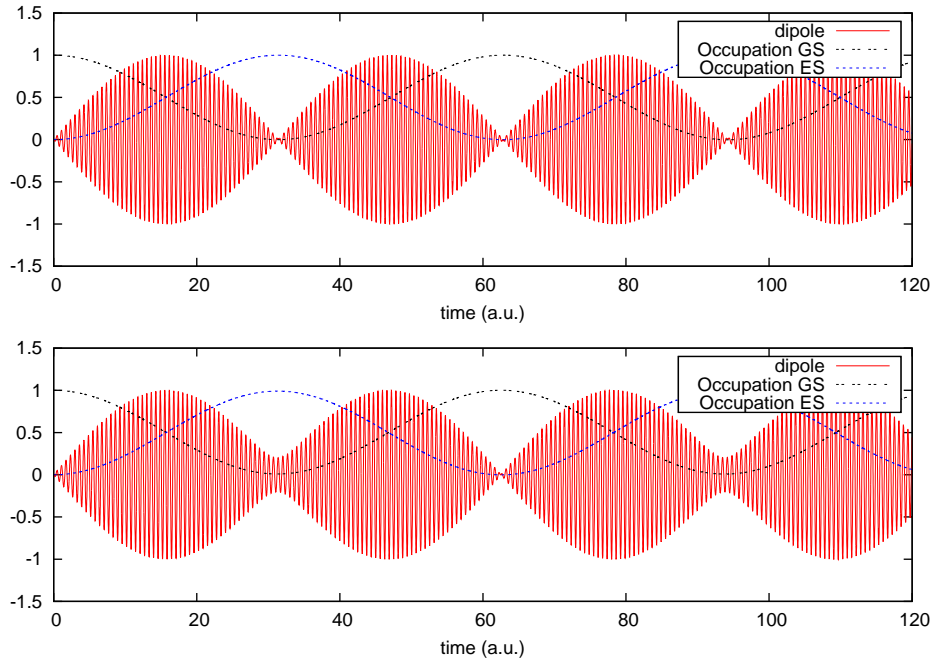}
    \caption{Time dependence of the expected value of the dipole and
      occupation number for the ground-state (GS) and exited state (ES)
      for exact tuning (upper panel) and a system with a frequency de-tuned by 0.01 (lower panel).}
  \label{fig:rabi}
\end{figure}

As before a gnuplot script is added in order to plot the
result Fig.\ref{fig:rabi}. 
This result is comparable to the one obtained in \cite{johRabi}.

\subsection{Collective Neutrino Oscillation}

In this example we implement a simple version of the collective
neutrino oscillation phenomena. We will follow the notation given
in \cite{raffelt}.

For a two level system and in the absence of non-coherent
interactions, the operators can be interpreted as a vectors in a three
dimensional space, in particular the state of the system is

\begin{equation}
\rho_w=P_w^i\lambda_i ,
\end{equation}
and the time independent Hamiltonian is
\begin{equation}
H_0=B_w^i\lambda_i
\end{equation}
where $i \in \{1,2,3\}$ and $w=T/E$, with $T$ being the temperature and $E$ the
energy of the state.

The evolution of the system is given by the density matrix equation

\begin{equation}
\dot \rho_w = i[H_0+\mu \rho,\rho_w] ,
\end{equation}
where $\rho=\int \rho_w dw$ and $\mu=\sqrt{2}G_F n_\nu$.

In order to get a geometrical description is useful to use the vector
notation: In the vector notation the commutator of the
operators is equivalent to the cross product of the vectors. Using this
the equation for a system with self interacting terms is given by

\begin{equation}
\dot P_w = (wB+\mu P) \times P_w,
\label{eq:Pw}
\end{equation}
where
\begin{equation}
P=\int P_w dw,
\label{eq:P}
\end{equation}
and  $\mu=\sqrt{2}G_F n_\nu$ is the self-interaction
strength, and B is the vacuum Hamiltonian. 

The problem we solve is the evolution of the density matrix for the
case where the parameter $\mu$ is varying until it reaches 0 at some time
$T$. As the initial condition we set all the vectors $P_w$ aligned
together, with a small angle $\theta$ from $B$.

\subsubsection{Derived object ({\ttfamily collective})}

The object is declared in \textrm{SQuIDS/examples/CollectiveNeutrinoOscillations/collective.h} 
and implemented in \textrm{SQuIDS/examples/CollectiveNeutrinoOscillations/collective.cpp}.

In this example, working in the interaction picture does not bring any
advantage for the case of having a large number of $w$ nodes. Therefore, we
do not need to define the evolved projectors.
Instead, and following the analogy of the vectors in three
dimensional space, we define the $SU(2)$ generators, which are equivalent the
unit vectors in the three perpendicular directions.

\begin{lstlisting}
  SU_vector ex,ey,ez;
  ex=SU_vector::Generator(nsun,1);
  ey=SU_vector::Generator(nsun,2);
  ez=SU_vector::Generator(nsun,3);
\end{lstlisting}

The constructor sets up the value of $\mu$ ({\ttfamily mu}), the angle between $B$ and
$P$ ({\ttfamily th}), the range for $w$ ({\ttfamily wmin} to {\ttfamily wmax}) and the number of $w$-bins ({\ttfamily Nbins}).
\begin{lstlisting}
  collective(double mu,double th, double wmin, double wmax, int Nbins);
  void init(double mu,double th, double wmin, double wmax, int Nbins);
\end{lstlisting}

Next we define

\begin{lstlisting}
void collective::PreDerive(double t){
  if(bar)
    progressbar(100*t/period, mu);
  //compute the sum of 'polarizations' of all nodes
  P=state[0].rho[0];
  for(int ei = 1; ei < nx; ei++)
    P+=state[ei].rho[0];
  //update the strength of self-interactions
  mu = mu_f+(mu_i-mu_f)*(1.0-t/period);
}
\end{lstlisting}
in which we implement Eq.\eqref{eq:P} and update the value of $\mu$.

We write {\ttfamily HI} to implement Eq.\eqref{eq:Pw}:
\begin{lstlisting}
SU_vector collective::HI(unsigned int ix,
                         unsigned int irho, double t) const{
  //the following is equivalent to
  //return Get_x(ix)*B+P*(mu*(w_max-w_min)/(double)nx);
  
  //make temporary vectors which use the preallocated buffers
  SU_vector t1(nsun,buf1.get());
  SU_vector t2(nsun,buf2.get());
  //evaluate the subexpressions into the temporaries
  t1=Get_x(ix)*B;
  t2=P*(mu*(w_max-w_min)/(double)nx);
  //return the sum of the temporaries, noting that t1 is no
  //longer needed so its storage may be overwritten
  return(std::move(t1)+t2);
}
\end{lstlisting}
Note that in the commented line we directly write Eq.\eqref{eq:Pw}, but for 
efficiency reasons we write equivalent code in a somewhat more verbose manner.
Instead of using unnamed temporary vectors we create temporary vectors using
preallocated memory buffers. This avoids the need to allocate memory on each
call to {\ttfamily HI}, although it comes at the cost that the function is neither reentrant 
nor thread-safe. In particular, to use this style one must ensure that no caller 
of {\ttfamily HI} ever keeps the result across a second call as it would be overwritten when
the buffer is reused.

The main function in the collective object is
\begin{lstlisting}
 void Adiabatic_mu(double mu_i, double mu_f, double period, bool bar); 
\end{lstlisting}
This function evolves the system, changing the parameter $\mu$ from $mu_i$
to $mu_f$ linearly during the period of time given by {\ttfamily period}. The
parameter {\ttfamily bar} controls whether the progress bar for the evolution is shown.

\subsubsection{Main file}

The main file declares two \textrm{collective} objects: one in order to
do the time evolution and the other as a reference.

\begin{lstlisting}[frame=leftline, numbers = left,breaklines=true]
int main(){
  //Parameters
  double mu=10.0;
  double mu2=10;
  double wmin=-2;
  double wmax=2;
  double th=0.01;
  double th2=0.01;
  int Nbins=200;
  
  collective ColNus(mu,th,wmin,wmax,Nbins);
  collective ColNus_notevolved(mu2,th2,wmin,wmax,Nbins);

  //Evolution from mu=10 to mu=0 in a time period of 100
  ColNus.Adiabatic_mu(10,0,100,true);

  SU_vector o=ColNus.ez;
  double max=0;
  //find the maximum of the initial spectrum
  for(int w=0;w<Nbins;w++){
    if(max<ColNus_notevolved.GetExpectationValue(o,0,w))
      max=ColNus_notevolved.GetExpectationValue(o,0,w);
  }

  //write the ouput in the file
  //col 1 value of w
  //col 2 expectation value of ez for the evolved system normalized to the maximum
  //col 3 expectation value of ez for the non evolved system normalized to the maximum
  std::ofstream file("collective.dat");
  for(int w=0;w<Nbins;w++){
    file << std::scientific << ColNus.Get_x(w) << "\t" 
	 << ColNus.GetExpectationValue(o,0,w)/max <<"  " <<  
      ColNus_notevolved.GetExpectationValue(o,0,w)/max << std::endl;
  }

  //runs the gnuplot script if yes
  std::string plt;
  std::cout << std::endl <<  "Done! " << std::endl <<  "Do you want to run the gnuplot script? yes/no" << std::endl;
  std::cin >> plt;
  
  if(plt=="yes" || plt=="y"){
    return system("./plot.plt");
  }
  
  return 0;
}
\end{lstlisting}

\begin{figure}[h]
  \centering
  \includegraphics{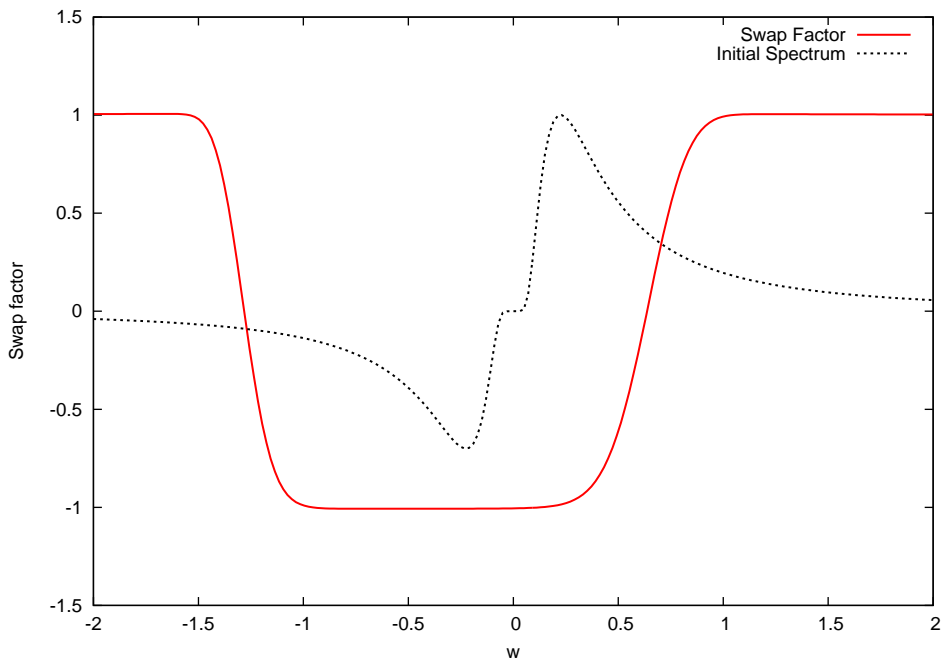}
    \caption{Swap factor (solid line) of the system after going from $\mu=10$ to
      $\mu=0$ in a time period of 100 time units. The dotted line
      shows the initial spectra.}
   \label{fig:collective_neutrino}
\end{figure}

After declaring the objects we evolve {\ttfamily ConNus} from $\mu=10$ to $\mu=0$
in a time period of $100$ time units. As before a gnuplot script is added in order to plot the
result, which is shown in Figure \ref{fig:collective_neutrino}.

{\bf Acknowledgments:} \ The authors acknowledge support from the Wisconsin IceCube Particle Astrophysics Center (WIPAC). C.A. and C.W. were supported in part by the
U.S. National Science Foundation under Grants No. OPP-0236449 and PHY-
0969061 and by the University of Wisconsin Research Committee with funds
granted by the Wisconsin Alumni Research Foundation.  
J.S. was funded under contract DE-FG-02-95ER40896 from the U. S. Department of Energy.

\bibliographystyle{elsarticle-harv_doi}

\bibliography{doc00}

\end{document}